\documentclass[aps,pre,twocolumn,nofootinbib,floatfix,showpacs]{revtex4}
\usepackage{epsfig}
\newcommand {\be}{\begin{equation}}
\newcommand {\ee}{\end{equation}}
\newcommand {\ba}{\begin{eqnarray}}
\newcommand {\ea}{\end{eqnarray}}

\begin{document}
%\draft
%\wideabs{

\title{Macroscopic detection of the strong stochasticity threshold in Fermi-Pasta-Ulam chains of oscillators}

\author{M. Romero-Bastida} 
\email{rbm@xanum.uam.mx}
\affiliation{Departmento de F\'\i sica, Universidad Aut\'onoma Metropolitana Iztapalapa, Apartado Postal 55--534, Distrito Federal 09340, M\'exico} 

\date{\today}

\begin{abstract}
The largest Lyapunov exponent of a system composed by a heavy impurity embedded in a chain of anharmonic nearest-neighbor Fermi-Pasta-Ulam oscillators is numerically computed for various values of the impurity mass $M$. A crossover between weak and strong chaos is obtained at the same value $\epsilon_{_T}$ of the energy density $\epsilon$ (energy per degree of freedom) for all the considered values of the impurity mass $M$. The threshold $\epsilon_{_T}$ coincides with the value of the energy density $\epsilon$ at which a change of scaling of the relaxation time of the momentum autocorrelation function of the impurity ocurrs and that was obtained in a previous work~[M. Romero-Bastida and E. Braun, Phys. Rev. E {\bf65}, 036228 (2002)]. The complete Lyapunov spectrum does not depend significantly on the impurity mass $M$. These results suggest that the impurity does not contribute significantly to the dynamical instability (chaos) of the chain and can be considered as a probe for the dynamics of the system to which the impurity is coupled. Finally, it is shown that the Kolmogorov-Sinai entropy of the chain has a crossover from weak to strong chaos at the same value of the energy density that the crossover value $\epsilon_{_T}$ of largest Lyapunov exponent. Implications of this result are discussed.
\end{abstract}
%}

\pacs{05.45.Jn, 05.45.Pq, 05.40.Jc}

\maketitle

%%%%%%%%%%%%%%%%%%%%%%%%%%%%%%%%%%%%%%%%%%%%%%%%%%%%%%%%%%
\section{Introduction}
%%%%%%%%%%%%%%%%%%%%%%%%%%%%%%%%%%%%%%%%%%%%%%%%%%%%%%%%%%

During the past two decades or so, there has been growing evidence of the connection between the underlying chaotic microscopic dynamics of many-particle systems and its observed macroscopic behavior. For example, the largest Lyapunov exponent (LLE) $\lambda_1$, which measures the exponential rate of divergence of two originally close trayectories in phase space, has been found to be an indicator (order parameter) of phase transitions~\cite{Butera,Mehra,Barre}. Moreover, the largest Lyapunov exponent has also been related to the transport coefficients of fluid systems with continuous potentials~\cite{Evans,Dorfman,Barnett}. For some special cases, e.g., hard-sphere systems, the theory of Lyapunov exponents is remarkably developed~\cite{Dorfman-book}. However, considering randomness and transport in general as signatures of microscopic chaos raises subtle and fundamental issues in statistical physics that have to be carefully discussed and investigated. For the case of Brownian motion the analysis of simplified models such as a labeled (test) particle immersed in a one-dimensional system of hard rods~\cite{Lebowitz} and an impurity in a harmonic crystal~\cite{Rubin}, has shown that the motion of the tracer particle may be Brownian even when the full dynamical system, particle plus fluid, is not chaotic. These examples show that microscopic chaos is sufficient, but not necessary to produce Brownian motion. The main problem then is to prove beyond doubt if it is possible to detect any particular feature of the microscopic dynamics, either regular or chaotic, in the behavior of the tracer particle. Only then it would be possible to assess the relevance of the microscopic dynamics on macroscopic behavior.

The FPU model, which is a one-dimensional chain of nearest-neighbor anharmonic oscillators, is a system that has been extensively studied over the past decades with relation to the problem of energy equipartition. From the dynamical system's perspective, it was the starting point in the study of chaotic dynamics in many-degrees-of-freedom systems (for a recent review, see Ref.~\cite{RevFPU}). In particular, numerical simulations~\cite{Pettini} revealed a rich phase space dynamics that is controlled by the energy per degree of freedom $\epsilon\equiv E/N$. Two qualitatively different regimes exist in the dynamical behavior of the system: it is strongly chaotic and phase-space diffusion (to be referred from now on as microscopic diffusion) is fast when the energy density $\epsilon$ exceeds a threshold $\epsilon_{_T}$, whereas it is only weakly chaotic (i.e., almost periodic) and microscopic diffusion is slowed down when the energy density is below the threshold $\epsilon_{_T}$, which in Ref.~\cite{Pettini} was called the {\sl strong stochasticity threshold} (SST). The detection of this transition between two different dynamical regimes is performed by means of the LLE $\lambda_1(\epsilon)$, which exhibits a change in its scaling behavior precisely at the value $\epsilon_{_T}$ of the SST.

Recently it was shown that, if the FPU chain is coupled to a heavy impurity, the latter performs Brownian motion~\cite{Romero}. This system, being one-dimensional, can be much easily studied than three-dimensional systems with continuous potentials, and offers a convenient starting point for a systematic study of the relationship between the microscopic dynamics and the macroscopic, statistical behavior. In particular, it is a suitable model to explore the possibility to find some indication of the known microscopic dynamics in the statistical behavior of the heavy impurity. The results of the present work suggest that this is the case indeed. In Sec. II we describe the model to be used. In Secs. III and IV we report the results of the statistical and dynamical behavior of the system, respectively. Section V is devoted to discuss the relationship between the results found in the two previous sections. In Sec. VI some conclusions are drawn.

%%%%%%%%%%%%%%%%%%%%%%%%%%%%%%%%%%%%%%%%%%%%%%%%%%%%%%%%%%
\section{The Model and its Numerical Investigation}
%%%%%%%%%%%%%%%%%%%%%%%%%%%%%%%%%%%%%%%%%%%%%%%%%%%%%%%%%%

In terms of dimensionless variables, the Hamiltonian of the model considered in this work is
\be
H\!=\!\!\!\!\!\sum_{i=-N/2}^{N/2} \displaystyle\left[\frac{p_i^2}{2m_i}+\frac{1}{2}(x_{i+1}-x_i )^2 +\frac{1}{4}\beta (x_{i+1}-x_i )^4 \right]~\label{newham}
\ee
\noindent
with $m_i=1$ for $i\not=0$ and $m_0=M$; periodic boundary conditions are assumed ($x_{(N/2)+1}=x_{-N/2}$). The model describes a system of one-dimensional $N$ coupled nonlinear oscillators of unit mass with nearest-neighbor interactions, displacements $\{x_i\}$, momenta $\{p_i\}$ and a central oscillator (impurity) of mass $M$ with displacement $x_0\equiv X$ and momentum $p_0\equiv P$. The value $\beta=0.1$ was used in the computation of all the numerical results hereafter reported. This model will be referred to as the Modified FPU (MFPU) model.

As initial conditions we choose the equilibrium value of the oscillators displacements, i.e. $x_i(0)=0$ for $i=-N/2,\ldots,N/2$. The momenta $\{p_i(0)\}$ were drawn from a Maxwell-Boltzmann distribution at temperature $T$ consistent with a given value of the energy density $\epsilon$. It is known that the dynamics of both the homogeneous (uniform mass, $M=1$) FPU model~\cite{Pettini} and the MFPU model~\cite{Romero} is strongly chaotic for large $\epsilon$ values, whereas their dynamics, for small $\epsilon$ values, corresponds to a chain of coupled harmonic oscillators. Consequently $\epsilon$ was chosen in the range $0.01\le\epsilon\le100$ with $M=1$, $40$, $60$, $80$, and $100$. Finally, the $2N$ first-order Hamilton equations of motion were integrated using a third-order bilateral symplectic algorithm~\cite{Casetti}, which is a high-precision numerical scheme.

%%%%%%%%%%%%%%%%%%%%%%%%%%%%%%%%%%%%%%%%%%%%%%%%%%%%%%%%%%
\section{Statistical Behavior of the Heavy Impurity}
\label{sec:SB}
%%%%%%%%%%%%%%%%%%%%%%%%%%%%%%%%%%%%%%%%%%%%%%%%%%%%%%%%%%

Thermal equilibrium between the impurity and the FPU chain with $N=300\,000$ unit mass oscillators is attained within the time interval of $t=5\times10^5$ natural time units. Afterwards the heavy impurity performs Brownian motion for all $\epsilon$ values studied~\cite{Romero}. The momentum autocorrelation function (MACF) $\rho_0(t)\equiv\langle P(t)P(0)\rangle_t/\langle P^2 (0)\rangle_t$ of the heavy impurity was obtained by computing the time averages $\langle\cdots\rangle_t$ over a time interval of $t=2\times10^5$. For all $\epsilon$ and $M$ values considered the exponential fit $\exp(-t/\tau)$, where $\tau$ is the relaxation time of the MACF, is valid for $t<50$. Since the magnitude of $\rho_0(t)$ is negligible for $t>50$ in all cases, its contribution was not considered in the computation of $\tau$.

\begin{figure}
\centerline{\epsfig{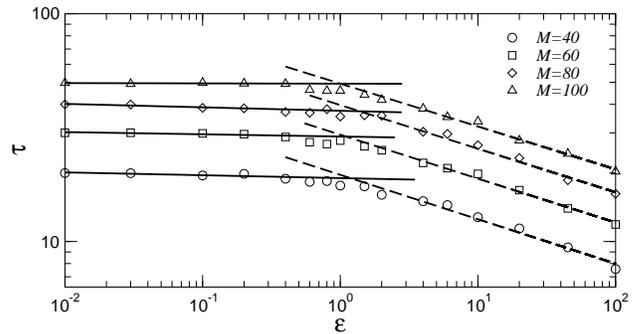}}
\caption{ Relaxation time $\tau$ vs energy density $\epsilon$. The continuous (dashed) lines correspond to the power-law fit $\epsilon^{\alpha}$ for the $\epsilon<1$ ($\epsilon\ge1$) regime and all $M\not=1$ values.}
\label{fig:Tauep}
\end{figure}

Fig.\ \ref{fig:Tauep} shows a graph of the relaxation time $\tau$ vs the energy density $\epsilon$ in log-log scale for all the considered $M$ values, as $\epsilon$ goes from the regular to the chaotic regime. We observe that the data points are separated into two different and well-defined regions, depending on the $\epsilon$ value. In all cases the dependence of $\tau$ on the energy density is weak when $\epsilon<1$. On the contrary, when $\epsilon\ge1$, $\tau$ decreases rapidly as $\epsilon$ increases. In each of these regimes $\tau$ has a power-law scaling $\tau_{_M}(\epsilon)=\tau_{_{0,M}}\epsilon^{\alpha_{_M}}$, with the same $\tau_{_{0,M}}$ value for all the data points with the same $M$ value. The slopes of each of the fits in Fig.\ \ref{fig:Tauep} are approximately the same for the $\epsilon<1$ regime. The same result occurs for $\epsilon\ge1$, though with another slope value. These facts imply that there is a common $M$-independent scaling exponent $\alpha$ for each regime. Following the methodology of Ref.~\cite{Romero} we estimate $\alpha\approx-0.019$ for $\epsilon<1$ and $\alpha\approx-0.182$ for $\epsilon\ge1$. Thus we can conclude that the power-law scaling exponent undergoes a sudden change at a threshold value $\epsilon_c\approx1.00$, since the estimated $\alpha$ values differ by one order of magnitude.

%%%%%%%%%%%%%%%%%%%%%%%%%%%%%%%%%%%%%%%%%%%%%%%%%%%%%%%%%%%%%%%%%%%
%%%%%%%%%%%%%%%%%%%%%%%%%%%%%%%%%%%%%%%%%%%%%%%%%%%%%%%%%%%%%%%%%%%

%%%%%%%%%%%%%%%%%%%%%%%%%%%%%%%%%%%%%%%%%%%%%%%%%%%%%%%%%%
\section{Phase-space dynamics}
\label{sec:DB}
%%%%%%%%%%%%%%%%%%%%%%%%%%%%%%%%%%%%%%%%%%%%%%%%%%%%%%%%%%

In Ref.~\cite{Romero} the change in the scaling behavior of $\tau$ depicted in Fig.\ \ref{fig:Tauep} was attributed to the change in the dynamical behavior of the chain as the energy density $\epsilon$ goes from a regime of low chaos to a regime of fully developed chaos. However, it remains unexplained why the aforementioned change occurs at a precise value $\epsilon_c$ of the energy density. To address this problem we consider in more detail the dynamical behavior of the FPU chain. As is known, the LLE $\lambda_1(\epsilon)$ is a parameter that can be used to quantify the degree of chaos in the dynamics of a given system with either low or large number of degrees of freedom. For the homogeneous FPU model~\cite{Pettini} it is known that the behavior of the LLE for large (small) values of the energy density $\epsilon$ is $\lambda_1(\epsilon)\sim\epsilon^{\sigma}$. The exponent $\sigma$ has a large value in the weakly chaotic regime $\epsilon<\epsilon_{_T}$ and a small value in the strongly chaotic regime $\epsilon>\epsilon_{_T}$. That is, the exponent $\sigma$ undergoes a sudden change around the threshold $\epsilon_{_T}$. This provides an operational definition of the SST, which is defined by a crossover in the scaling behavior of $\lambda_1(\epsilon)$. Further analytical studies~\cite{Casetti-geometry,Ruffo} of the dependence of the LLE on the energy density $\epsilon$ in homogeneous FPU lattices confirm the conjecture that the LLE reaches a finite, $\epsilon$-dependent, value in the thermodynamic limit $N\rightarrow\infty$.

\begin{figure}
\centerline{\epsfig{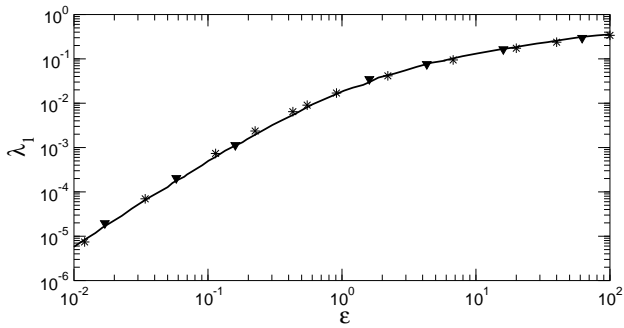}}
\caption{ Largest Lyapunov exponent $\lambda_1$ vs energy density $\epsilon$ for the homogeneous FPU model ($M=1$). Asterisks correspond to $N=2000$ and filled triangles to $N=300\,000$. The continuous line is the analytical estimate of Ref.~\cite{Casetti-geometry}.}
\label{fig:LambdaM1}
\end{figure}

For the homogeneous FPU model we have numerically computed the LLE by a standard technique~\cite{Benettin} for chains with $N=300\,000$ and $N=2000$ unit mass oscillators, as shown in Fig.\ \ref{fig:LambdaM1}. The results for these two $N$ values overlap very well over the entire range of the $\epsilon$ values studied with the best available analytical estimate of $\lambda_1(\epsilon)$ valid in the thermodynamic limit~\cite{Casetti-geometry}. Since our numerical data are independent of the approximations made in Ref.~\cite{Casetti-geometry} to obtain the analytical estimate, they provide an independent corroboration of the stability of $\lambda_1(\epsilon)$ in the thermodynamic limit. This result justifies the study of the dynamics of the FPU chain with $N$ values smaller than those needed to compute the relaxation time $\tau$ of the MACF of the heavy impurity.

%%%%%%%%%%%%%%%%%%%%%%%%%%%%%%%%%%%%%%%%%%%%%%%%%%%%%%%%%%%%%%%%%%%
%%%%%%%%%%%%%%%%%%%%%%%%%%%%%%%%%%%%%%%%%%%%%%%%%%%%%%%%%%%%%%%%%%%

%%%%%%%%%%%%%%%%%%%%%%%%%%%%%%%%%%%%%%%%%%%%%%%%%%%%%%%%%%%%%%%%%%%
\begin{figure}
\centerline{\epsfig{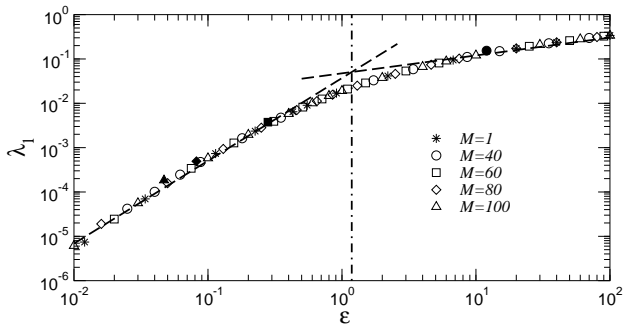}}
\caption{ Largest Lyapunov exponent $\lambda_1$ vs energy density $\epsilon$ for the MFPU model and all $M$ values. References to power-law scalings valid for small and large $\epsilon$ values are shown by dashed lines. Vertical dashed-dotted line indicates the approximate location of $\epsilon_{_T}$. Open symbols correspond to $N=2000$ and filled symbols to $N=300\,000$.}
\label{fig:Lambda}
\end{figure}
%%%%%%%%%%%%%%%%%%%%%%%%%%%%%%%%%%%%%%%%%%%%%%%%%%%%%%%%%%%%%%%%%%%

Fig.\ \ref{fig:Lambda} shows the results for the MFPU model with $N=2000$ unit mass oscillators. For all the $M$ values considered the scaling behavior of $\lambda_1$ remains unaltered by the presence of the heavy impurity embedded in the chain for the entire $\epsilon$ value range considered. This result indicates that the dynamics of the system is dominated by the FPU chain, with no contribution from the heavy impurity. A rough estimate of the crossover energy density, which defines a common SST $\epsilon_{_T}$ for both the homogeneous FPU and MFPU models, can be obtained from Fig.\ \ref{fig:Lambda} by the intersection of the straight lines that describe the low and high energy power-law asymptotic behavior of $\lambda_1(\epsilon)$. The crossover between the two power laws ocurrs at $\epsilon_{_T}\approx1.18$. A straightforward observation is that the computed SST $\epsilon_{_T}$ value, which is a distinctive characteristic of the microscopic dynamics of the entire system, is very close to the value $\epsilon_c\approx1.00$ obtained from the crossover in the scaling behavior of the relaxation time $\tau$ of the MACF, which is a property of the heavy impurity. This result suggests that the change in the scaling behavior of $\tau$ exhibits macroscopically the existence of the SST $\epsilon_{_T}$.

\begin{figure}
\centerline{\epsfig{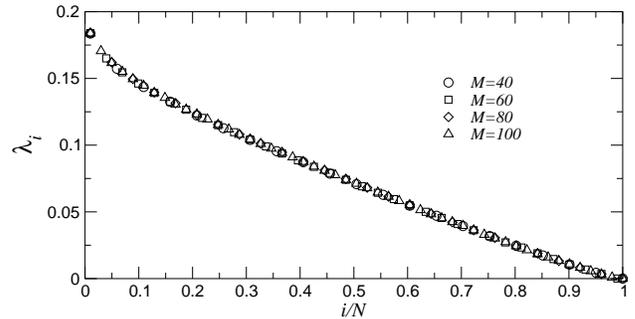}}
\caption{ Spectrum of Lyapunov exponents for $\epsilon=26$, $N=100$, and all the considered $M\not=1$ values.}
\label{fig:Lesp} 
\end{figure}

To characterize chaos in more detail we compute the Lyapunov spectrum (LS) of positive exponents $\{\lambda_i \}$ using a standard method~\cite{Giorgilli}, as implemented in Ref.~\cite{Wolf}, for chains of $N=100$ unit mass oscillators and several $M$ values. The results are presented in Fig.\ \ref{fig:Lesp}. The linear shape of the LS has been previously obtained for the homogeneous FPU model with $\epsilon$ values much larger than the SST $\epsilon_{_T}$~\cite{Livi}. We see that the LS is unaltered by the heavy impurity; that is, the LS is dominated by the positive Lyapunov exponents corresponding to the FPU chain without the impurity. This type of behavior has been reported for the case of a heavy tracer particle in a two-dimensional molecular fluid, where the tracer and fluid particles are hard-disks undergoing elastic collisions~\cite{Gaspard}, but not for the case of a continuous potential, like the one considered in this work.

%%%%%%%%%%%%%%%%%%%%%%%%%%%%%%%%%%%%%%%%%%%%%%%%%%%%%%%%%%%%%%%%%%%
%%%%%%%%%%%%%%%%%%%%%%%%%%%%%%%%%%%%%%%%%%%%%%%%%%%%%%%%%%%%%%%%%%%

\begin{figure}
\centerline{\epsfig{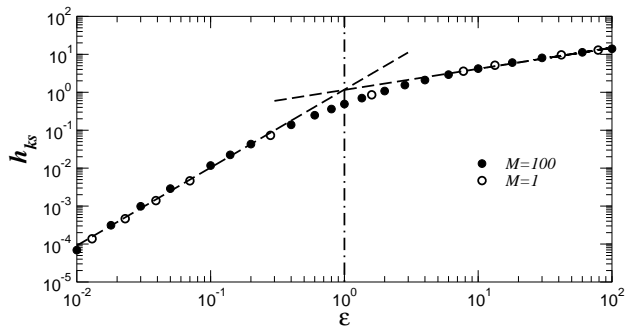}}
\caption{ KS entropy vs energy density for $N=100$. Vertical dashed-dotted line indicates the approximate crossover point between the regimes of weak and strong chaos, with dashed lines indicating the asymptotic power-laws valid for $\epsilon\ll1$ and $\epsilon\gg1$.}
\label{fig:Hks} 
\end{figure}

Another useful quantity to characterize chaos is the Kolmogorov-Sinai (KS) entropy $h_{_{KS}}$ which, for conservative systems, can be written as $h_{_{KS}}=\sum\lambda_i$ for all $\lambda_i\ge0$. The KS entropy describes the mean information production rate caused by all positive Lyapunov exponents along a trajectory in phase-space and therefore measures the degree of stochasticity. Results for a chain with $N=100$ and two $M$ values are reported in Fig.\ \ref{fig:Hks}. We observe that the KS entropy, just as the LLE, exhibits a change in its scaling behavior between two different and well-defined regimes, characterized by the scaling laws $h_{_{KS}}\sim\epsilon^{2.05}$ for $\epsilon\ll1$ and $h_{_{KS}}\sim\epsilon^{0.55}$ for $\epsilon\gg1$. The crossover between both regimes occurs at an energy density value $\epsilon$ that is closer to the threshold value $\epsilon_c$ of the relaxation time $\tau$ than the threshold $\epsilon_{_T}$ of the LLE. Thus the KS entropy, just as the LLE, is a suitable quantity to probe the dynamics in the considered $\epsilon$ value range. A similar behavior in the KS entropy has been previously reported for the one dimensional $\phi^4$ model, with a threshold value of the energy that separates two integrable limits of this system~\cite{Mutschke}.

%%%%%%%%%%%%%%%%%%%%%%%%%%%%%%%%%%%%%%%%%%%%%%%%%%%%%%%%%%%%%%%%%%%
%%%%%%%%%%%%%%%%%%%%%%%%%%%%%%%%%%%%%%%%%%%%%%%%%%%%%%%%%%%%%%%%%%%

%%%%%%%%%%%%%%%%%%%%%%%%%%%%%%%%%%%%%%%%%%%%%%%%%%%%%%%%%%
\section{Discussion}
%%%%%%%%%%%%%%%%%%%%%%%%%%%%%%%%%%%%%%%%%%%%%%%%%%%%%%%%%%

From a macroscopic, i.e. statistical, perspective it is not at all clear what kind of collective mechanism, if any, could be held responsible for the crossover value $\epsilon_c$ at which the scaling exponent $\alpha$ changes its value. However, from a microscopic Hamiltonian perspective the change of the scaling exponent $\alpha$ can be straightforwardly interpreted as a macroscopic manifestation of the chaotic transition described by the SST. This interpretation arises by comparing Figs.\ \ref{fig:Tauep} and\ \ref{fig:Lambda}, which suggest that the behavior of $\tau_{_M}(\epsilon)$ and $\lambda_1(\epsilon)$ is due to the same mechanism. Since the microscopic dynamics of the MFPU and FPU models are practically identical, we can apply the same explanation of the origin of the chaotic transition that was previoulsy given in context of the FPU model~\cite{Pettini}. The Hamiltonian of either the FPU or the MFPU model can be written in the form
\be
H(\theta,{\bf I})=H_{_0}({\bf I})+H_{_1}(\theta,{\bf I}),\qquad\mu\equiv\frac{||H_{_1}||}{||H_{_0}||}\ll1,
\ee
where $(\theta,{\bf I})$ are the action-angle canonically conjugated variables and $||\cdots||$ is a suitable norm. A consequence of the perturbation $H_{_1}$ is that the resonant manifolds ${\bf n}\cdot\omega({\bf I})=0$ of $H_{_0}$ are destroyed for any small $\mu$ and are replaced by finite-thickness chaotic layers (${\bf n}$ is an integer component vector and $\omega$ is a vector whose components are $\omega_i=\partial H_{_0}/\partial I_i$). As this chaotic surfaces intersect the constant energy hypersurface for $N\gg1$, a chaotic network (the Arnold web) is produced which is everywhere dense in phase space. At $\epsilon>\epsilon_c$ the resonances are strongly overlapped and microscopic diffusion is allowed in every direction in phase space, with a large value of the LLE. In this dynamical regime the scaling exponent $\alpha$ has also a large value, which produces a rapid decay in the value of the relaxation time $\tau$ of the MACF of the heavy impurity. On the contrary, at $\epsilon<\epsilon_c$, the resonance overlapping is drastically reduced as the energy density decreases, diffusion in phase space occurs only along resonances, and the LLE takes a smaller value. Correspondingly, the scaling exponent $\alpha$ is smaller, which accounts for the almost constant value of the relaxation time $\tau$ in this $\epsilon$ value range. The behavior of the relaxation time $\tau(\epsilon)$ has a direct relationship with diffusion in configuration space, since $\tau(\epsilon)$ is directly related to the self-diffusion coefficient of the heavy impurity through the Green-Kubo relation~\cite{Romero}

Although it is not our aim to directly address the problem of the origin of diffusion in many-particle Hamiltonian systems, a discussion will clarify some of our results. Recently a class of one-dimensional maps has been reported which present normal diffusive-like behavior in the absence of chaos~\cite{Cecconi}. For the case of the MFPU model the diffusion coefficient of the heavy impurity obeys a unique power-law $D\sim\epsilon^{0.964}$ in the entire range of $\epsilon$ values considered~\cite{Romero}. From these results it is clear that macroscopic diffusion is a combined result of the large number of degrees of freedom, the mass of the heavy impurity~\cite{NoteMass}, and the random initial conditions, rather than of the degree of chaos in the system. However, although diffusion itself is a collective phenomenon and, therefore, largely independent of the microscopic dynamics, some of its {\sl specific} features, such as the crossover in the scaling behavior of $\tau$, can yield information on the dynamics of the system to which the heavy impurity is coupled, which in the case of the FPU chain studied here is the transition from weak- to a strongly chaotic regime. Nevertheless, in order to consider the heavy impurity as an effective probe of the Hamiltonian dynamics we have to study a suitable macroscopic variable, that in our case is the relaxation time $\tau$, which is computed from the momentum or velocity of the heavy impurity.

Finally we would want to remark some implications of the behavior of the KS entropy $h_{_{KS}}$. Although $\tau$ is a property computed from the momentum of the heavy impurity, whereas the KS entropy is essentialy a property computed from the full dynamics of the FPU chain, both quantities provide information about the microscopic dynamics. In fact, both reflect the chaotic transition at the SST $\epsilon_{_T}$, as shown in Figs.\ \ref{fig:Lambda} and\ \ref{fig:Hks}. The important point to be stressed is that the KS entropy is a property that can be computed by other methods, such as time series-analysis~\cite{Kantz}. Recent theoretical work suggests that the information extracted from the time-series of the position of a Brownian particle is unable to unambiguously determine the nature (either regular, chaotic, or stochastic) of the system to which the particle is coupled~\cite{Cencini}. This result is entirely consistent with the lack of any signature of the chaotic transition associated to the SST in the behavior of the self-diffusion coefficient as a function of $\epsilon$ in the case of our MFPU model~\cite{Romero}. In Ref.~\cite{Cencini} the KS entropy of an impurity in a harmonic crystal, which corresponds to our MFPU model for $\epsilon\ll1$, was computed using the time-series of the position of the impurity. However, if the KS entropy is computed using the full dynamics of the system, as was done in the present paper for the case of the MFPU model, it can indeed detect the stochasticity transition, as is evident by inspecting Fig.\ \ref{fig:Hks}. By comparing Figs.\ \ref{fig:Lambda} and\ \ref{fig:Hks} it can be infered that $\lambda_1$ and $h_{_{KS}}$ convey almost the same information, which in turn is reflected in the behavior of the relaxation time $\tau(\epsilon)$ as depicted in Fig.\ \ref{fig:Tauep}. Now, $\tau(\epsilon)$ is computed from the MACF, which is a property that depends only on the momentum of the impurity. This observation suggests that, if the momentum time-series, instead of the position time-series, is used to compute the KS entropy by methods of time-series analysis, the stochasticity transition associated with the SST of the FPU model could indeed be detected. We stress that we are not implying that methods based on time-series analysis can detect chaos in a generic dynamical system. Even for one-dimensional systems, this possibility remains controvertial~\cite{Cencini}. The dynamics of three-dimensional systems or those with long-range interactions is still not well characterized, which poses an obstacle when trying to define which specific features of the microscopic dynamics of these type of systems could be detected at a macroscopic level. But, for the particular case of the MFPU model, it can be conjectured that the KS entropy, when computed from the momentum time-series, would display a different behavior at small $\epsilon$ values compared to the corresponding behavior at large $\epsilon$ values.

%%%%%%%%%%%%%%%%%%%%%%%%%%%%%%%%%%%%%%%%%%%%%%%%%%%%%%%%%%%%%%%%%%%
\section{Conclusions}
%%%%%%%%%%%%%%%%%%%%%%%%%%%%%%%%%%%%%%%%%%%%%%%%%%%%%%%%%%%%%%%%%%%

In this work we have presented evidence, obtained from a systematic study of certain dynamical parameters, that the inclusion of a heavy defect in a FPU chain does not affect its Hamiltonian dynamics. Furthermore, the crossover in the scaling behavior of both the LLE and the KS entropy, which depend on the dynamics of the whole system, has a macrospcopic manifestation in a similar behavior of the relaxation time $\tau$ of the MACF of the heavy impurity alone. An interesting development in this direction would be to explore the possibility that certain dynamical features could be detected in more complicated systems by monitoring the apropiate variables. Another open problem is to investigate if this transition between two diferent dynamical regimes can be detected by other methods, such as those provied by nonlinear time-series analysis.

\begin{acknowledgments}
I wish to acknowledge M.~A.~Nu\~nez and Professor L.~Garc\'\i a-Col\'\i n for their comments and suggestions. I am grateful to L. Casetti for letting me have his numerical code, which has been an invaluable starting point for the development of my own. Financial support from CONACyT, M\'exico is also acknowledged.
\end{acknowledgments}

%%%%%%%%%%%%%%%%%%%%%%%%%%%%%%%%%%%%%%%%%%%%%%%%%%%%%%%%%%%%%%%%%%%%%%%%%%%%

\end{document}